\documentclass[12pt,preprint]{emulateapj}

\def\be{\begin{equation}}
\def\ee{\end{equation}}
\def\bea{\begin{eqnarray}}
\def\eea{\end{eqnarray}}
\usepackage{graphicx}
\usepackage{amsmath}
\usepackage[colorlinks,
            linkcolor=blue,
            anchorcolor=blue,
            citecolor=blue]{hyperref}
\usepackage{color,txfonts}
\begin{document}

\title{Measure the Propagation of a halo CME and Its Driven Shock with 
the Observations from a Single Perspective at Earth}

\author{Lei Lu\altaffilmark{1,2,4}, Bernd Inhester\altaffilmark{2}, Li Feng\altaffilmark{1,3}, Siming Liu\altaffilmark{1}, Xinhua Zhao\altaffilmark{3}}
\email{E-mail:inhester@mps.mpg.de (BI); lfeng@pmo.ac.cn (LF)}

\altaffiltext{1}{Key laboratory of Dark Matter and Space Astronomy, Purple Mountain Observatory, Chinese Academy of Sciences, Nanjing 210008, China}
\altaffiltext{2}{Max Planck Institute for Solar System Research, Justus-von-Liebig-Weg 3, 37077 G\"{o}ttingen, Germany}
\altaffiltext{3}{State Key Laboratory of Space Weather, National Space Science Center, Chinese Academy of Sciences, Beijing 100190, China. }
\altaffiltext{4}{University of Chinese Academy of Sciences, Yuquan Road 19, Beijing, 100049, China}

\begin{abstract}
We present a detailed study of an earth-directed coronal mass ejection (Full halo CME) 
event happened on 2011 February 15 making use of white light observations by three 
coronagraphs and  radio observations by Wind/WAVES. We applied three different methods 
to reconstruct the propagation direction and traveling distance of the CME and its driven shock. 
We measured the kinematics of the CME leading edge from white light images observed by 
STEREO A and B, tracked the CME-driven shock using the frequency drift observed by Wind/WAVES 
together with an interplanetary density model, and obtained the equivalent scattering centers of 
the CME by Polarization Ratio(PR) method. For the first time, we applied PR method to different 
features distinguished from LASCO/C2 polarimetric observations and calculated their projections 
onto white light images observed by STEREO A and B. By combining the graduated cylindrical shell 
(GCS) forward modeling with the PR method, we proposed a new GCS-PR method to derive 3D parameters 
of a CME observed from a single perspective at Earth. Comparisons between different methods show 
a good degree of consistence in the derived 3D results.
\end{abstract}


\setlength{\parindent}{.25in}

\section{Introduction}
Coronal Mass Ejections (CMEs) are powerful eruptions that release huge clouds 
of plasma threaded with magnetic field lines from the Sun into interplanetary space.
These events have an influence on the entire near-Sun heliosphere. 
 On Earth they can cause technical problems to power grids, 
oil pipelines, or telecommunication equipments \citep{Pirjola02}. Moreover, high 
energy particles propelled from these events could be a potential threat to human 
life in space\citep{Facius06}. In order to predict the arrival of a CME and avoid potential 
damages, many methods have been developed to monitor the position and 
three-dimensional (3D) structure of CMEs using data from multiple coronagraphs. 
These methods include stereoscopy \citep{Inhester06,Aschwanden08}, GCS Forward
Modeling (GCSFM) 
\citep{Thernisien06}, Polarization Ratio method (PR) \citep{Moran04}, mask fitting \citep{Feng12}, 
and the local correlation tracking plus triangulation \citep{Mierla09}. Comparisons 
between different methods have been made by \cite{Mierla10}, \cite{Thernisien11} 
and \cite{Feng13}.

Of particular importance to space weather are the so-called halo CMEs. They 
propagate in direction close to the Sun-Earth line and have been observed 
by coronagraphs on board different near-Earth spacecraft such as P78-1, 
SMM and the SOHO \citep{Domingo95}. In 2006 the twin 
STEREO spacecraft were launched to monitor transient events in interplanetary 
space from two vantage points off the Sun-Earth line, and to determine their 3D 
locations \citep{Kaiser08}. The observations from this mission have greatly
improved the determination of CME propagation.
However, the lifetime of the STEREO mission is limited 
and currently STEREO-B has only been recovered after an almost two year loss of
contact. The state of the instrument is not yet clear. It is therefore of
some interest to find out how well we can predict the 
propagation direction and speed of a halo CME from observations 
made from a single perspective alone, especially from a near-Earth position.
Most future missions equipped with coronagraphs, e.g., ESA's 
PROBA-3\citep{Zhukov14}, Indian Aditya-L1\citep{Nandi15}, and Chinese ASO-S\citep{Gan15}, 
will image the inner corona from about 1.1 to 3 $R_\odot$ from the near-Earth
perspective. The other two future missions which will escape Earth, Solar Orbiter and
Solar Probe Plus \citep{Velli13}, will have highly ecliptic orbits
not well suited for a synoptic CME watch.
It is therefore not clear to what extent and with which precision we will in
the future be able to routinely determine the propagation characteristics of
CMEs based on near-Earth observations alone.

The goal of the present paper is
to test two methods to analyze halo CME propagation which are 
independent of the STEREO position geometry. The STEREO data will be
used in this paper as a reference to find out how reliable these alternative
methods are.
One such alternative are characteristic radio
burst signals, called type II bursts, usually generated in association with CMEs.
Previous studies have shown that type II radio bursts in the decameter to
hectometer (DH) 
and longer wavelength range are produced by CME-driven shocks
\citep{Reiner98,Bale99,Su16} and 
bear signatures which reflect the propagation of the transient 
events through the interplanetary space (IP space). The observed frequency drift rate 
can be converted into an approximate velocity of the CME and its shock if a model for the
upstream electron density with distance from the Sun is assumed. 

However, the source region of the type II radio bursts is still an open question. 
\cite{Gopalswamy04} suggested that the type II radio bursts are enhanced and modified 
due to the interaction between two CMEs. \cite{Oliveros12} applied the radio direction-finding 
technique to an interacting CME event, and compared the results with the white light 
observations by STEREO from which they concluded that type II radio emission is causally 
related to the interaction of CMEs. \cite{Magdalenic14} applied the same technique to another 
event, and suggested that the interaction between the shock wave and a nearby coronal 
streamer resulted in the type II radio emission, which is consistent with the conclusion 
of \cite{Shen13} .

The PR technique is a second alternative. It was first proposed by \cite{Moran04} to convert 
the polarimetric observations by LASCO/C2 to 3D distances off the plane of the sky (POS), 
and verified by \cite{Dere05} using a series of high-cadence (1 hr) LASCO polarization 
measurements. Later on, \cite{Mierla09} and \cite{Moran10} successfully applied this technique to the polarimetric 
observations from STEREO coronagraphs. The physics behind this method is Thomson 
scattering and it has been described in detail by \cite{Billings66} and reviewed by \cite{Howard09} 
and \cite{Inhester15}. Like the observation of type II radio bursts, the method has the advantage 
that only the observations from one single perspective is required. Previous studies usually apply 
this method to limb CMEs where foreground and background contamination of the CMEs is 
not as severe as for the case of halo CMEs. As halo CMEs are most relevant to geomagnetic 
storms, we have made efforts to obtain their 3D location with this method by 
carefully removing other, CME-irrelevant features in the polarimetric observations. For most of 
the halo CMEs, especially for full halo CMEs, we are lacking the CME observations behind the 
coronagraph occulter. In order to obtain the 3D CME structure as completely as possible, 
we also applied the graduated cylindrical shell (GCS) model to fit the halo
CME observations and the corresponding 3D points derived with PR method.

In this paper, we select a full halo CME as seen by SOHO. The two STEREO spacecraft 
were almost in a opposite direction from the Sun and made an angle of about $90^\circ$ with SOHO. 
Such a geometry of spacecraft positions simplifies the stereoscopy method and is favorable 
for comparing the results derived from different methods. In Section~\ref{sec-obs} we describe 
the instruments and observations. In Section~\ref{sec-rec} we present details of analyses 
methods and their corresponding results. The comparison of the results derived from different 
methods are shown in Section~\ref{sec-com}. Section~\ref{sec-sum} gives a
summary of our work.

\section{Observations}\label{sec-obs}

The CME investigated here was observed on 15 February 2011 from three
viewpoints almost simultaneously by the coronagraphs on board the two Solar
Terrestrial Relations Observatory (STEREO) \citep{Kaiser08} probes and the
Solar and Heliospheric Observatory (SOHO) \citep{Domingo95} spacecraft. Each 
of the twin STEREO probes is equipped apart from other instruments with two white-light 
coronagraphs COR1 and COR2. Their field of view ranges from 1.5 to 4 solar radii 
and from 2 to 15 solar radii, respectively. The SOHO has two white-light 
coronagraphs, C2 and C3, on board which together image the solar corona from 
2.2 to 30 solar radii(C2: 2.2--6 solar radii, C3: 3.7--30 solar radii)\citep{Brueckner95}. 
All of the telescopes can provide polarized and total brightness images.

\begin{center}
\begin{figure*}[!htb]
\centering
\includegraphics[width=0.8\textwidth]{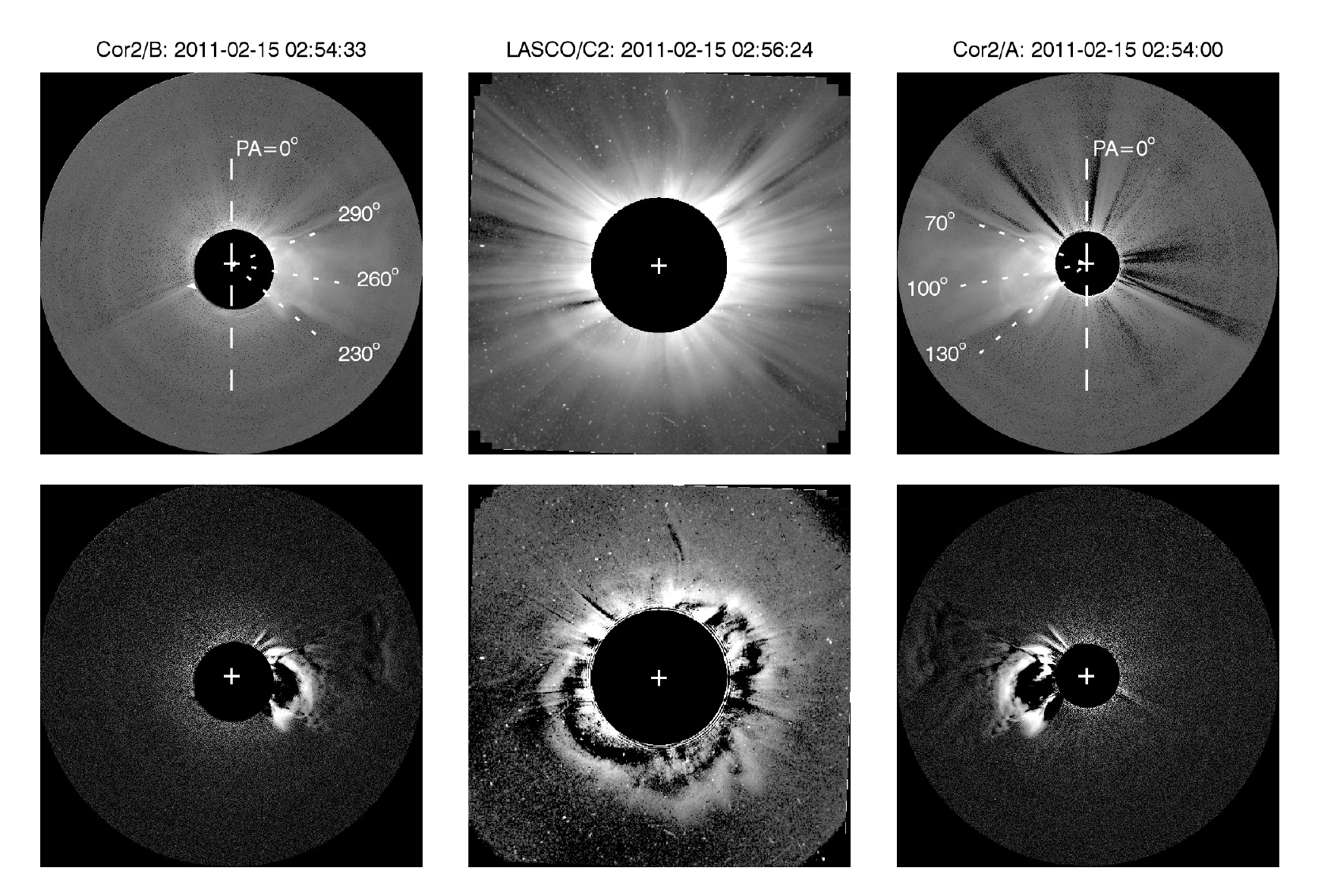}
\caption{The panels (from left to right) show the image triplets  recorded by STEREO-B/COR2 
    at 02:54:33, by SOHO/LASCO C2 at 02:56:24, and by STEREO-A/COR2 at 02:54:00 on 
   15 February 2011, respectively. The upper panels show the images with minimum backgrounds 
   subtracted, the lower panels show the corresponding running-difference images.  The dotted 
   lines in the upper row indicate different position angles.}
\label{F-wl}
\end{figure*}
\end{center}

During the period of 13-15 February 2011, there were eight CMEs ejected from the active 
region AR 11158 as pointed by \citet{Maricic14}. Most of these CMEs were too weak
to trace their shape with the desired precision. Therefore, we focus our
analysis on a more intense full halo
CME, which was first captured  by STEREO/COR1 at 01:55UT on 15 
February 2011, and was associated with an X2.2 flare. Another reason for us to select 
this event is the special viewing geometry of the three spacecraft, which gave a full  
view of the CME from different perspectives. The positions of the three spacecraft 
during this event are presented in Table \ref{tab-pos}. The view directions of the 
STEREO spacecraft almost form a right angle with the view direction of SOHO. 
Due to this special geometry of the three spacecraft, the CMEs recorded as a halo 
CME by SOHO was observed as a limb event by both STEREO A and B.
Figure~\ref{F-wl} shows  the image triplets  recorded by STEREO-B/COR2 
 at 02:54:33, by SOHO/LASCO C2 at 02:56:24, and by STEREO-A/COR2 at 02:54:00 on 
 15 February 2011, respectively. The upper panels show the images with minimum backgrounds
 subtracted (monthly minimum background for STEREO images and two-day minimum 
 background for LASCO/C2 image. The chosen of two-day minimum background for C2 will be 
 explained in detail in Section \ref{subsubsec-method}), the lower panels show the corresponding 
 running-difference images.

\begin{table}[!htb]
\centering
\caption {Longitude and latitude of STEREO-A/B and SOHO in the Carrington coordinate system on 15 February 2011.}
\begin{tabular}{cccc}
\hline \hline
 Spacecraft& STEREO & SOHO & STEREO-A \\
\hline
 Longitude($^\circ$) & -92.64 & 21.20 & 108.16\\
 Latitude($^\circ$) & 3.22 & -6.82 & -2.80 \\
 Angular separation STEREO A - B  &  & 179.10 &  \\
\hline
\hline
\end{tabular}
\label{tab-pos}
\end{table}

\begin{center}
\begin{figure*}[!htb]
\centering
\includegraphics[width=0.8\textwidth]{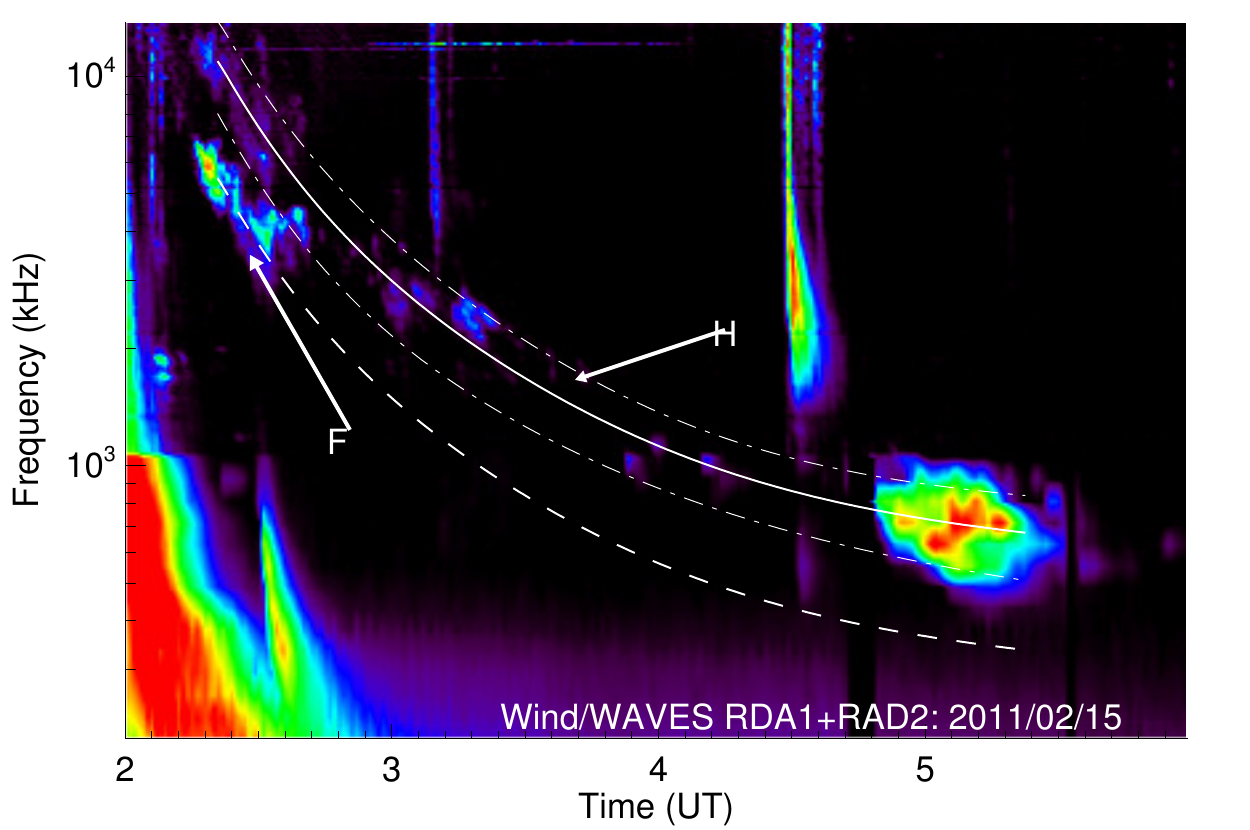}
\caption{The spectrogram(400kHz--13.825MHz) observed by Wind/WAVES 
receivers RAD1/RAD2. The dashed and solid lines represent the fundamental and its 
second harmonic plasma emission, respectively. The dot-dash lines indicate the  boundaries 
of the second harmonic emission.  All of these lines are fitted with the interplanetary 
density model by \cite{Vrsnak04}.}
\label{F-spectrogram}
\end{figure*}
\end{center}

The radio emission associated with the CME-driven shock started at about 02:00UT, and
was observed by two SWAVES instruments on board STEREO~\citep{Kaiser08} and
the WAVES instrument on board the WIND spacecraft~\citep{Bougeret95}. 
Figure~\ref{F-spectrogram} shows the strong intermittent type II burst in decameter 
to kilometer range recorded by WIND/Waves. From these recordings we distinguish two lanes 
which correspond to the fundamental plasma emission and its second harmonic. 

\section{Reconstructions}\label{sec-rec}
In this Section we describe the processing of the data and
the shape and distance reconstruction of the CME based on it.
A comparison of the individual results will follow in Section~\ref{sec-com}.

\subsection{Reconstruction from stereoscopy}\label{subsec-ste}

Coronagraph images only give the two-dimensional (2D) projection of CMEs onto the
plane-of-sky (POS) normal to the respective view direction. Therefore, the speed obtained 
from the projected distance of the apparent CME leading edge is typically underestimated 
if the CME propagation direction is off the POS. For the halo CME investigated here, the 
propagation direction does not lie far from the POS of STEREO A and B. On the
other hand, a genuine stereoscopic reconstruction is not straight forward, because 
STEREO A and B have almost opposite view directions, hence their image information is 
almost redundant while the view from SOHO does not allow to discern the leading edge.
Instead it rather yields the lateral extent of the CME.

\begin{center}
\begin{figure*}[h]
\centering
\includegraphics[width=0.8\textwidth]{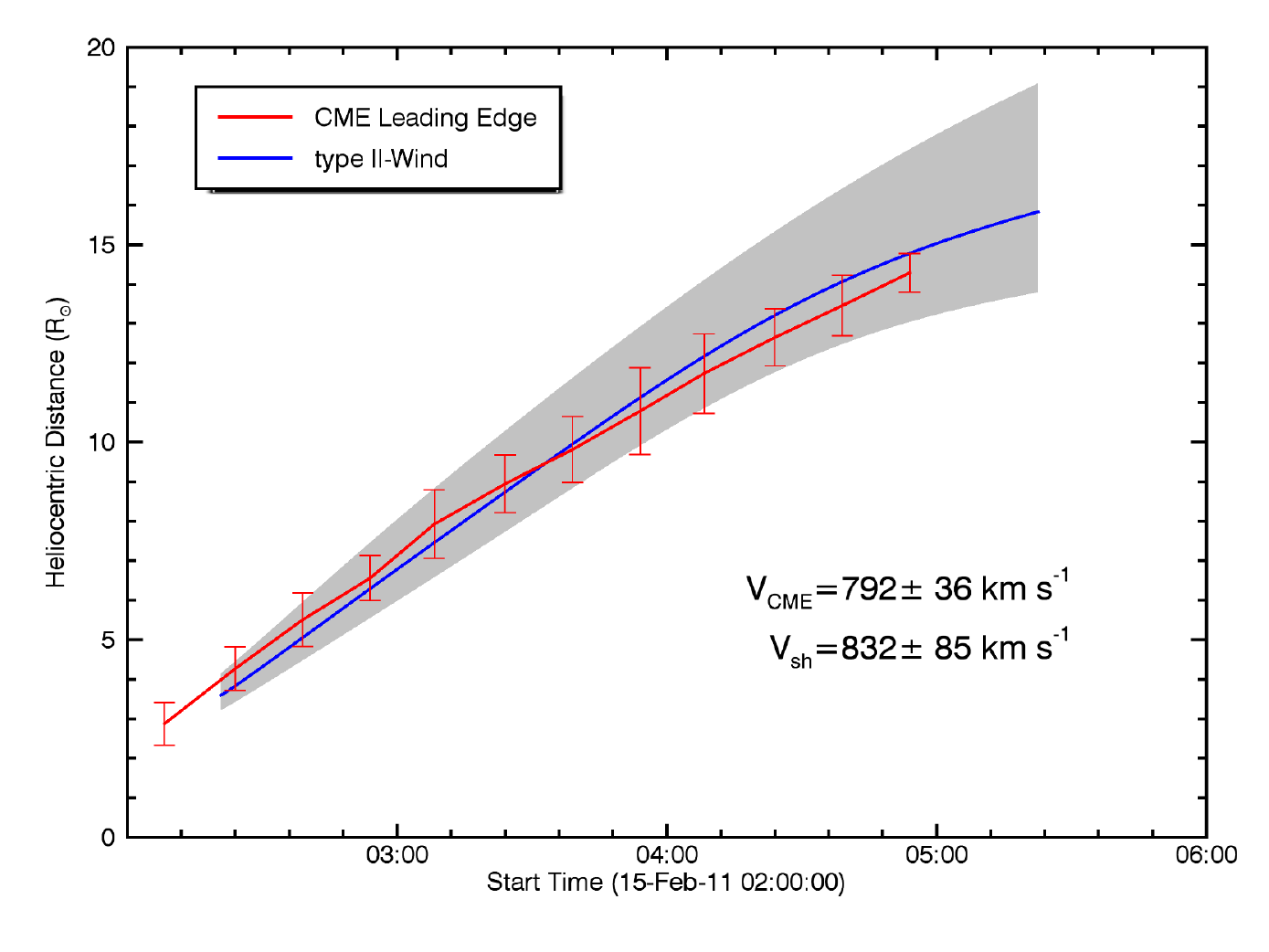}
\caption{Kinematics of the projected leading edge of the CME and the source region of the 
   associated type II radio burst. The leading edge was measured from STEREO observations(red line). 
   The radio source region was derived from the observations of the type II radio burst using the model proposed
   by \cite{Vrsnak04}(blue line). The red vertical bars indicate the uncertainties in identifying the leading 
   edge, the grey region shows the position uncertainties of the radio source region. The velocity of the CME 
   was estimated to be $\rm 792 \pm 36~km~s^{-1}$ and the velocity of the radio source was estimated to be 
   $\rm 832 \pm 85 ~ km~s^{-1}$.}
\label{F-th}
\end{figure*}
\end{center}

 Using the data of STEREO/COR2 A and B, we measure the kinematics of the projected 
 leading edge of the CME along different position angles (PAs) as indicated
 in Figure~\ref{F-wl}. The red line in Figure~\ref{F-th} shows the measured distance-time profile 
 for the CME leading edge and the red vertical bars indicate the distance uncertainties  
 which are due to the kinematic dependence on the PAs and the intensity drop off when identifying 
 the outermost bright CME structure.
The top and the bottom of the bars indicate 
 the maximum and minimum heights of the CME leading edge from the Sun center along 
 different PAs. The average propagation speed of the leading edge was estimated to be 
 $\rm 792 \pm 36 ~{km~s^{-1}}$ by linearly fitting the
 distance-time profiles along different PAs recorded by both STEREO/COR2 A and B.        

\begin{center}
\begin{figure*}[h]
\centering
\includegraphics[width=0.8\textwidth]{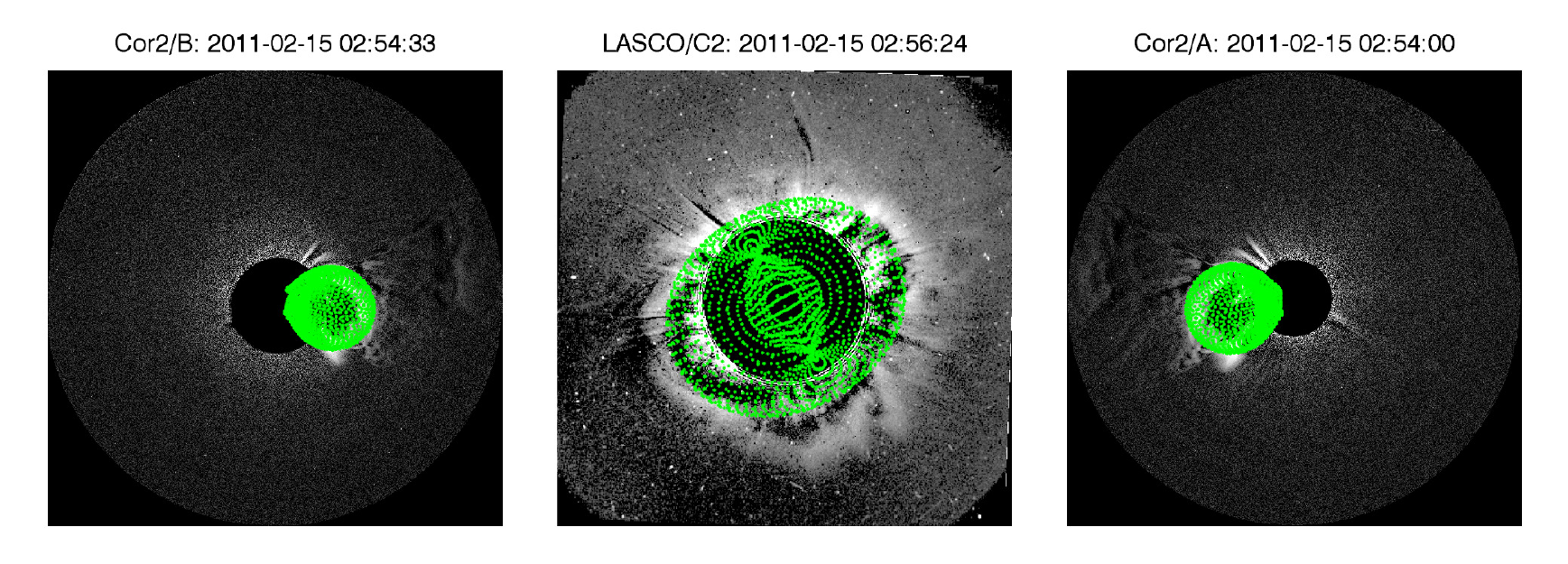}
\caption{The same image triplets as shown in the lower row in Figure~\ref{F-wl}. For this time, the wireframe from the 
   graduated cylindrical shell(GCS) model was overlaid on top, marked by green mesh 
   points for each image.}
\label{F-gcs}
\end{figure*}
\end{center}

As an alternative approach to stereoscopy, a fit of the visible CME boundaries
in several coronagraph images with a parameterized flux rope CME model
(GCSFM) has become popular (\citealp{Thernisien06},\citeyear{Thernisien09}).
While both methods rely on stereoscopy, the GCSFM fits the CME shape interactively to
a family of flux-rope-like surfaces of six geometrical parameters,
the line-tying approach does not make any a-priori assumptions about the
CME shape but is often restricted to its leading-edge surface.
We apply GCSFM to our data at the time when LASCO provided the
polarized image set as shown in Figure~\ref{F-gcs}, from  which  we estimate that at about 02:54 UT
the CME propagated to a height of $\rm 7.1 R_\odot$ and its center was directed at a longitude 
of $22^\circ$ and latitude of $-9^\circ$ in a carrington coordinate system. 
These values are also listed in Table~\ref{tab-cme}.

\subsection{Reconstruction from frequency inversion}\label{subsec-rad}

The emission of Type II bursts is generally assumed to be produced by
electron density fluctuations generated by energetic electron beams.
The acceleration process is
not known in detail, but there is strong evidence that it occurs in the
vicinity of the shock which runs ahead of the CME front. If the CME front
propagates within the solar wind frame at superalfv\'enic speed, the stand-off
distance between CME front and the shock ahead should not change too much
\citep{Reiner98}.
The instability of the
electron beam generates electrostatic and by linear coupling also
electromagnetic noise at the plasma frequency and its harmonics.
The relation between the plasma frequency and the background electron density
is \citep{Priest82}:
\begin{equation}  \label{density-frequency} 
  N_{\rm e}=(\frac{f_{\rm pe}[\rm Hz]}{8.98\times10^3})^2\quad \rm cm^{-3}
\end{equation}
The observation of the frequency with time can therefore be converted
to a distance vs time estimate of the emission region from the Sun if an
interplanetary plasma density model is assumed.

There is a variety of interplanetary 
density models, which were developed based on different measurements. 
Discrepancies become obvious when matching the IP densities with Active 
Region(AR) corona. For example, the model by \citet{Leblanc98}, which agrees 
well with in-situ observations by the Hellios spacecraft, gives too low densities 
when close to the Sun surface, especially when compared with the AR corona. On 
the other hand, the model by \citet{Saito70}, being very successful when applied 
to the AR corona, gives too high densities at large distances in IP space. In our 
analysis, we fit the visible parts of the type II burst to the model by \citet{Vrsnak04}, 
who used the relation  $B \propto \frac{1}{R^2}$ between
 the corona magnetic field $B$ and the height $R$ to 
 smoothly connect the AR corona and IP space.
 The density model was normalized to the electron density at 1 AU of
 $\rm N_e(1AU)=3.46~cm^{-3}$, the average value observed by Wind/SWE 
 before CMEs arrived.   
The normalized density model is given by
\begin{equation} \label{density-model}
\begin{split}
N_{\rm e}(r)=&1.59\times10^5 r^{-2}+4.81\times 10^7 r^{-4}+1.52\times 10^8 r^{-6} 
                       \\
                      & +7.42\times 10^8 r^{-16}~ \rm{cm^{-3}}
\end{split}
\end{equation}
where $r$ is the distance in unit of $\rm R_\odot$, $\rm 1AU=215 R_\odot$.

To obtain smooth frequency drifts for the type II burst in the dynamic spectrum, we 
propose the following function according to the adopted density model, 
 \begin{eqnarray} \label{fitting-function}
(f_{\rm pe})^2=at^{-2}+bt^{-4}+ct^{-6}+dt^{-16}
\end{eqnarray}
where a,b,c,d are fit parameters which incorporate the still unknown speed $v$
and its possible time variation.

 In Figure~\ref{F-spectrogram} we present our fitting results\footnote{Since the second harmonic lasts over 
 a much longer time than the fundamental, we pick up a set of points from the radio spectrogram for the 
 second harmonic branch and fitted them with the model given by Equation (3).}.
 The dashed and solid lines represent the fundamental and its second harmonic 
 plasma emission, respectively. For the the second harmonic emission we also fit
 the visible boundaries indicated by the dot-dashed lines. The difference
 between the dot-dashed lines will be used to 
 estimate the relative position uncertainties of the radio source region. In our analysis, the second 
 harmonic band was used to estimate the heliocentric distance of the source. The 
 corresponding results are presented in Figure~\ref{F-th} (blue line). The gray region indicates 
 the resulting uncertainty in the height of the radio source region. The average
 velocity of the radio source is estimated to be $\sim \rm 815~km~s^{-1}$, which
 is slightly faster but still
 comparable to the speed determined form the leading edge of the CME seen in
 the STEREO images. The position of the radio source at 02:56:24 when SOHO/LASCO 
 took the polarized image sequence is estimated to be between $\rm 5.7 R_\odot$ and 
 $\rm 7.7 R_\odot$.

\subsection{Reconstruction from the Polarisation Ratio(PR) method}\label{subsec-pol}

\subsubsection{Method description}\label{subsubsec-method}
The polarization of sunlight scattered by corona electrons is well
known \citep{Billings66}.
The scattering cross section depends on the angle between the scattering direction and 
the electric field vector. Since the light emitted from the photosphere is 
unpolarized, it can be split into two equal, mutually perpendicular components,
one normal to and one in the scattering plane.
From Thomson Scattering theory, the scattered intensity of the former (denoted as $I_{tan}$) 
is independent of the scattering angle $\chi$, while the scattered intensity of
the latter (denoted 
as $I_{rad}$)  varies as $\sin^2\chi$. The polarized brightness $I_{pol}$, total brightness $I_{tot}$ 
and polarization degree P are defined as 

\begin{eqnarray}
  I_\mathrm{pol}&=&I_\mathrm{tan}-I_\mathrm{rad}\\
  I_\mathrm{tot}&=&I_\mathrm{tan}+I_\mathrm{rad}=2I_\mathrm{tan}-I_\mathrm{pol}\\
  P&=&\frac{I_\mathrm{pol}}{I_\mathrm{tot}}
\end{eqnarray}

\begin{figure}[h]    
   \plotone{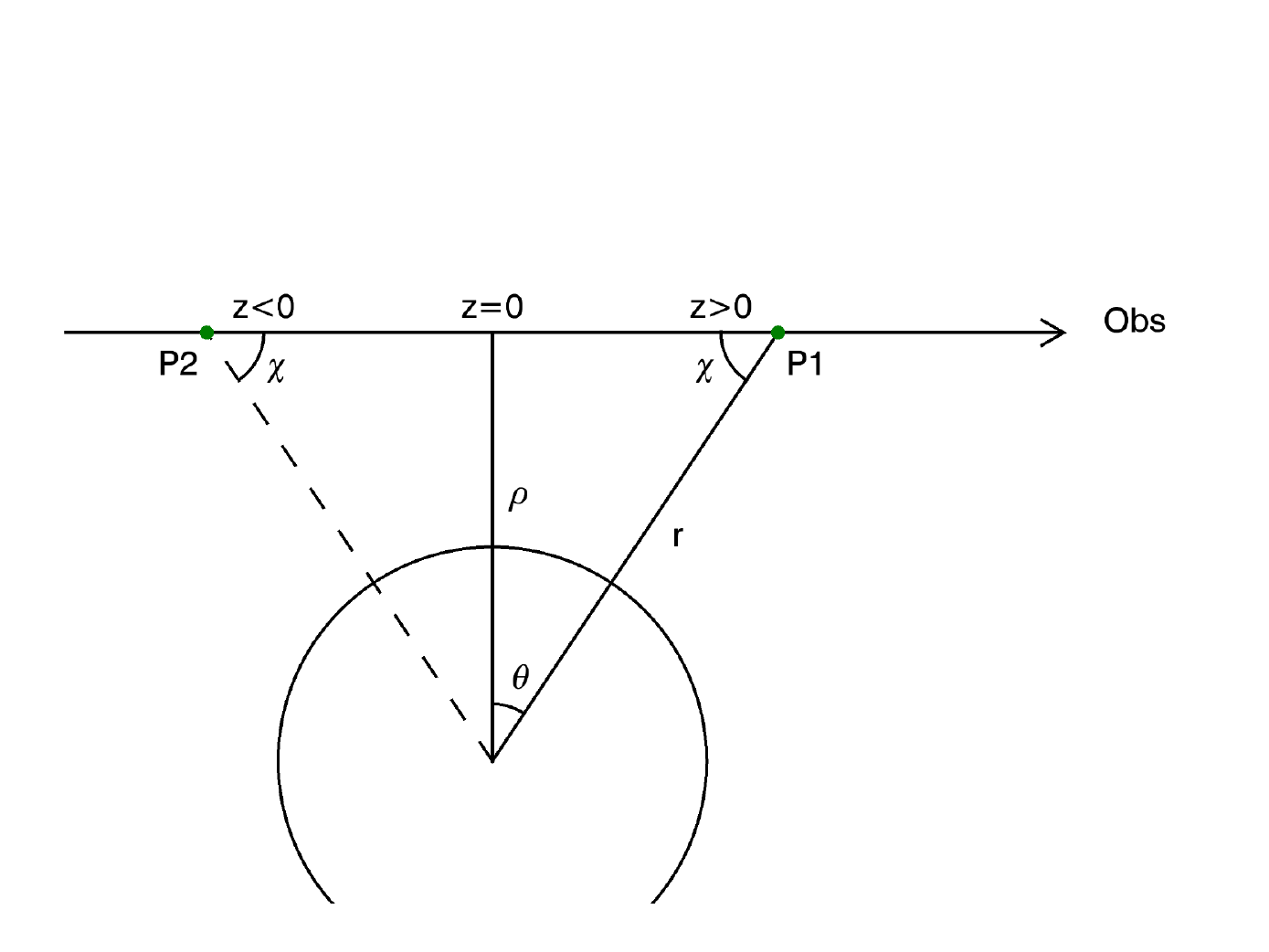}
   \caption{A sketch illustrating the geometry of Thomson scattering along one line-of-sight. 
   $\chi$ is scattering angle at position P1, z is the distance from the POS, 
   $\rho$ is the projected distance along the POS, r is the heliocentric distance. 
   P2 is the symmetrical position of P1. \label{F-variable_def}}
 \end{figure}

The expressions of $I_{tan}$ and $I_{pol}$ were given by, e.g., \citet{Howard09}:
\begin{eqnarray}
  I_\mathrm{tan}&=&\frac{\pi\sigma_e}{2}I_{\circ}\int_{0}^\infty dz\;N_e(\rho,z) [(1-u)C+uD] \label{eq-tan1}\\
  I_\mathrm{pol}&=&\frac{\pi\sigma_e}{2}I_{\circ}\int_{0}^\infty dz\;N_e(\rho,z)[(1-u)A+uB]\sin^2\chi \label{eq-pol1}
  \end{eqnarray}
 where $I_\circ$ is the intensity of solar disk center; $\sigma_e$ is the Thomson scattering 
cross section; u is the limb-darkening coefficient; $N_e$ is the local electron density; 
$\rho$ is the projected distance along the POS; z is the distance along the line-of-sight 
from POS;   and A, B, C, D could be expressed as functions of $\Omega$, the half-angle subtended by the solar disk at the scattering point, 
\begin{eqnarray}
  A(r)&=&\cos\Omega(r)\sin^2\Omega(r)\\
  B(r)&=&-\frac{1}{8}\left[1-3\sin^2\Omega(r)-\frac{\cos^2\Omega(r)}{\sin\Omega(r)}(1+3\sin^2\Omega(r)) \right.
                \nonumber  \\
                &&    \qquad \qquad
               \left. \ln\left(1+\frac{\sin\Omega(r)}{\cos\Omega(r)}\right)\right] \\
  C(r)&=&\frac{4}{3}-\cos\Omega(r)-\frac{1}{3}\cos^3\Omega(r) \\
 D(r)&=&\frac{1}{8}\left[5+\sin^2\Omega(r)-\frac{\cos^2\Omega(r)}{\sin\Omega(r)}(5-\sin^2\Omega(r)) \right.
            \nonumber \\
            && \qquad \qquad
            \left. \ln(\frac{1+\sin\Omega(r)}{\cos\Omega(r)})\right] 
 \end{eqnarray}
  where the angle $\Omega$ is given by  
 $\sin \Omega(r)=1/r $~(r is the heliocentric distance in unit of solar radii~$R_\odot$,~$r^2=\rho^2+z^2$).
 In Figure~\ref{F-variable_def}, we illustrate 
the geometry of these variables.

For each line of sight, a conventional assumption of the PR method is that all the electrons, which contribute to $I_{tan}$ and $I_{pol}$, 
are located at one single position($\rho_\circ$, $z_\circ$) (indicated as P1 in Figure~\ref{F-variable_def}) which we refer to as the equivalent scattering center. The corresponding electron density is assumed to be $N_e(\rho_\circ, z_\circ)$, then 
Equations \ref{eq-tan1} and \ref{eq-pol1} can be converted to

 \begin{eqnarray}
  I_\mathrm{tan}&=&\frac{\pi\sigma_e}{2}I_{\circ}N_e(\rho_\circ,z_\circ) [(1-u)C+uD] \label{eq-tan2}\\
  I_\mathrm{pol}&=&\frac{\pi\sigma_e}{2}I_{\circ}N_e(\rho_\circ,z_\circ)[(1-u)A+uB]\sin^2\chi \label{eq-pol2}
  \end{eqnarray}
Under this assumption the polarization degree can be expressed as the ratio of
polarized to the total brightness
\begin{equation}
P=\frac{[(1-u)A+uB]\sin^2\chi}{2[(1-u)C+uD]-[(1-u)A+uB]\sin^2\chi}  \label{eq-polarization}
\end{equation}
where A,B,C,D and $\chi$ are functions of $\rho_\circ$, $z_\circ$.

From Equation \ref{eq-polarization}, the theoretically derived relationship between 
polarization degree P, projected distance $\rho_\circ$ and distance from POS $z_\circ$ for the 
equivalent scattering center is shown in the left panel of Figure~\ref{F-pdeg}. On the other hand, $P$ 
and $\rho_\circ$ can be obtained from observations and therefore it's possible to get some 
estimate of the line-of-sight distance of the scatterers by solving (\ref{eq-polarization})
for $z_\circ$.

However, the method has some drawbacks:
The observed scattered signal is the result of a line-of-sight integration.
Depending on the electron density distribution along the line-of-sight,
the observed polarization degree is influenced from a wide distance range
while the formal application of the relation in the left panel in Figure~\ref{F-pdeg} just yields
a single distance.
Moreover, (\ref{eq-polarization}) depends on the square $z_\circ^2$ so that
the sign of the distance of the equivalent
scattering center from the plane-of-sky is ambiguous unless
the context or measurements from another view point allow to distinguish
whether the scatterer is in front or behind the plane-of-sky.
For example, in Figure~\ref{F-variable_def} points P1, P2 are equivalent and
yield the same polarization ratio \citep{Dai14}.  
On the day when the halo CME occurred, SOHO/LASCO C2 unfortunately produced only one 
polarized image sequence every six hours, and the CME was imaged in
polarization mode only at a single instance at 02:56:54UT.
We therefore have only one single polarized image set which we can use to 
compare the PR method with the distances estimated from the previous two 
methods.

In order to remove the background from the images and isolate the scattering
of the CME alone, we need to subtract background images. Usually 
two kinds of methods for background subtraction are used. One method is to subtract the 
pre-event images which are taken just before the event, and the other method is to 
subtract minimum images build from the minimum value of each pixel over all the images 
during a specific time range (usually one month, which is  long enough to subtract the F-corona and the stray light). In our case, the 
pre-event images were taken too early to remove the background well enough and a monthly minimum background leaves 
too many streamers so that the CME could not be well identified, therefore we 
produce a two-day minimum background from the polarized images taken on 14-15 February 2011 for each of the three polarizers
at $\rm -60^\circ$, $\rm 0 ^\circ$, $\rm 60^\circ$.
We subtracted the respective background from the three primary polarized brightness 
images rather than to subtract a background from the archived pB image.
because the polarized brightness is non-linearly related to the primary measurements, 
It therefore makes a difference to the conventional approach to subtract the background 
from the archived ready-made pB images. Naming $I_{\phi^\circ}$ the polarized
brightness at polarization angle $\phi$ with the respective background
subtracted, the polarized brightness $pB_{\rm obs}$, total brightness $tB_{\rm obs}$
and polarization degree $P_{\rm obs}$ of the CME are determined from\citep{Billings66}

\begin{eqnarray}
          pB_{\rm obs}&=&\frac{4}{3}[(I_{-60^\circ}+I_{0^\circ}+I_{60^\circ})^2-  
          \nonumber \\
          &&
           3(I_{-60^\circ}I_{0^\circ}+I_{-60^\circ}I_{60^\circ}+I_{0^\circ}I_{60^\circ})]^{1/2}  \\
          tB_{\rm obs}&=&\frac{2}{3}(I_{-60^\circ}+I_{0^\circ}+I_{60^\circ})    \\
          P_{\rm obs}&=&\frac{pB_{\rm obs}}{tB_{\rm obs}}
\end{eqnarray}

Before calculating the polarization degree, we run a 3-by-3-pixel averaging box over the 
synthesized pB and tB images to increase the signal-to-noise ratio and reduce the 
possible error caused by the CME motion during the exposures of the three polarizers. 
In Figure~\ref{F-features} we show our analyses of the LASCO polarimetric 
observations for the large-scale quiet Sun in upper panels and for the investigated CME 
in lower panels. The quiet Sun polarimetric observations were taken at 02:56:30 on 14 February 2011 
when no CMEs appeared in the field of view(FOV) of C2. A monthly minimum 
background was subtracted from each polarized image before synthesizing tB and pB. 
Quite a number of streamer structures can be clearly distinguished in the latitude range from 
about -60 to 60 degrees. Our analyses indicate that streamers have relatively larger polarization degree and smaller 
distances from POS than the ambient corona, which are consistent with the theoretical predictions. 
The lower panels display the polarimetric observations of the halo CME 
taken at 02:56:24 on 15 February 2011. Unlike the quiet Sun, a two-day minimum background was subtracted from 
each polarized image to show the CME structure as clearly as possible. Details of the structures
appearing in the polarimetric images will be presented in the following Section.

\subsubsection{Different features in C2 images} \label{subsec-fea}
\begin{center}
\begin{figure*}[!htb]
\centering
\includegraphics[width=0.8\textwidth]{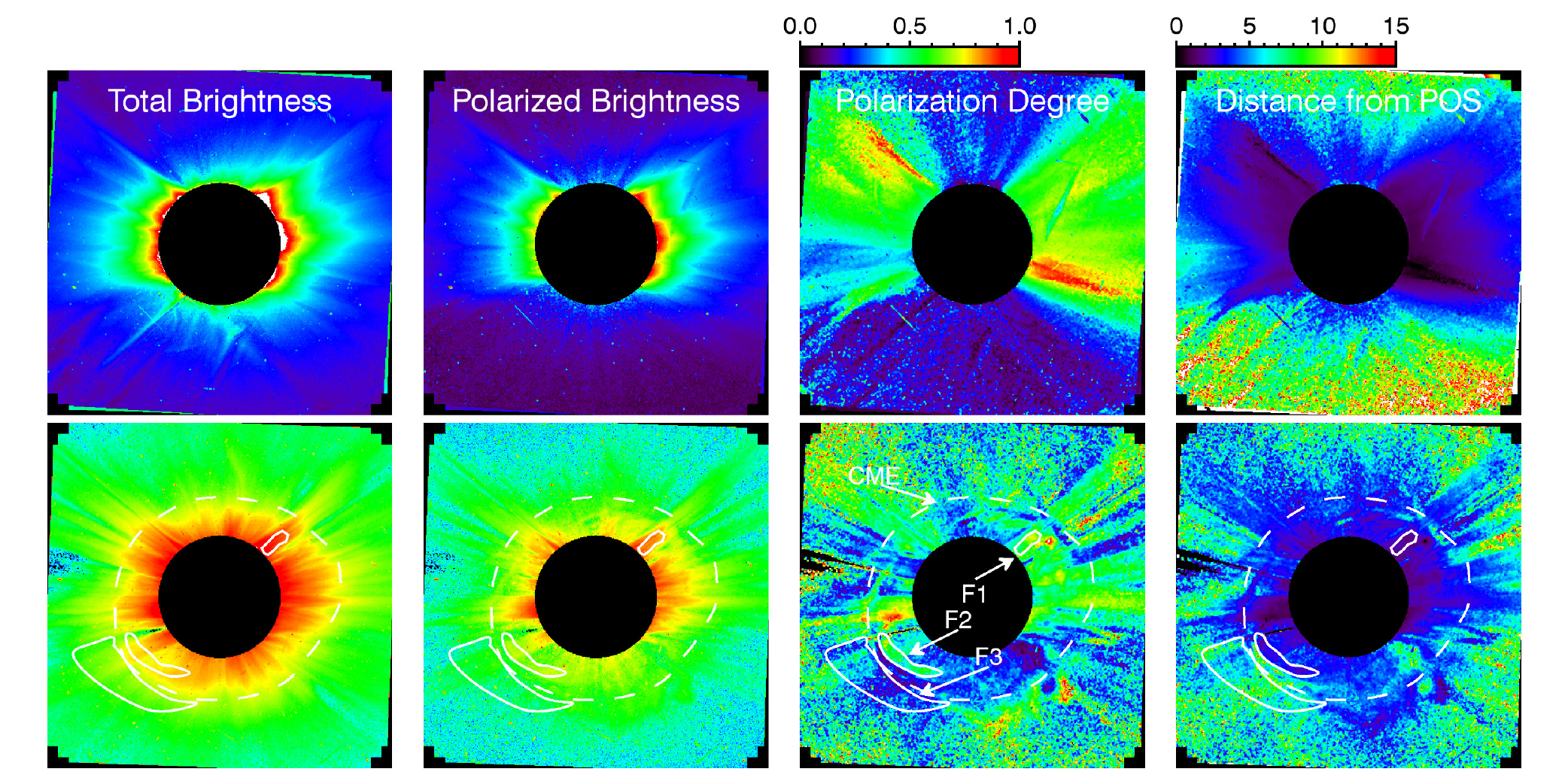}
\caption{Optical polarimetric observations from LASCO/C2. The upper row shows the 
   observations of a large-scale quiet Sun with monthly minimum background subtracted. The measurements 
   were made at 02:56:30 on 14 February 2011. The lower row shows
   the observations of the halo CME with two-day minimum background subtracted. The measurements were made at 02:56:24 on 15 February 2011. The panels in 
   each row (from left to right) show the total brightness, the polarized brightness, the distribution of 
   polarization degree and the distribution of distances(in unit of solar radii) from Plane of Sky, respectively. 
    The dashed curves in the lower row indicate the boundary 
    of the CME identified from white-light observations, and the solid curves indicate 
    the boundaries of three different features(marked by F1, F2, F3) distinguished from the image of polarization degree.}
\label{F-features}
\end{figure*}
\end{center}

The image of the polarization degree of the halo CME in Figure~\ref{F-features} shows a wealth of
coronal structures which were superposed by the line-of-sight integration. The
color code represents the polarization ratio and helps to separate
features from different depths.
We can distinguish bright background streamer structures which
have not been eliminated by the two-day minimum background subtraction.
To single out the halo CME in the foreground, we try to identify and discard
the near-Sun coronal background based on the polarization ratio.
We have marked some of the presumable background features with enhanced polarization
(marked by F1, F2, F3)in Figure~\ref{F-features}.

F1 has a high polarization and extends radially from the occulter 
edge to 2.6 $\rm R_\odot$. The polarization and shape suggest that the feature 
can be attributed to a streamer in the background corona, not far from the POS. 

\begin{center}
\begin{figure*}[!htb]
\centering
\includegraphics[width=0.8\textwidth]{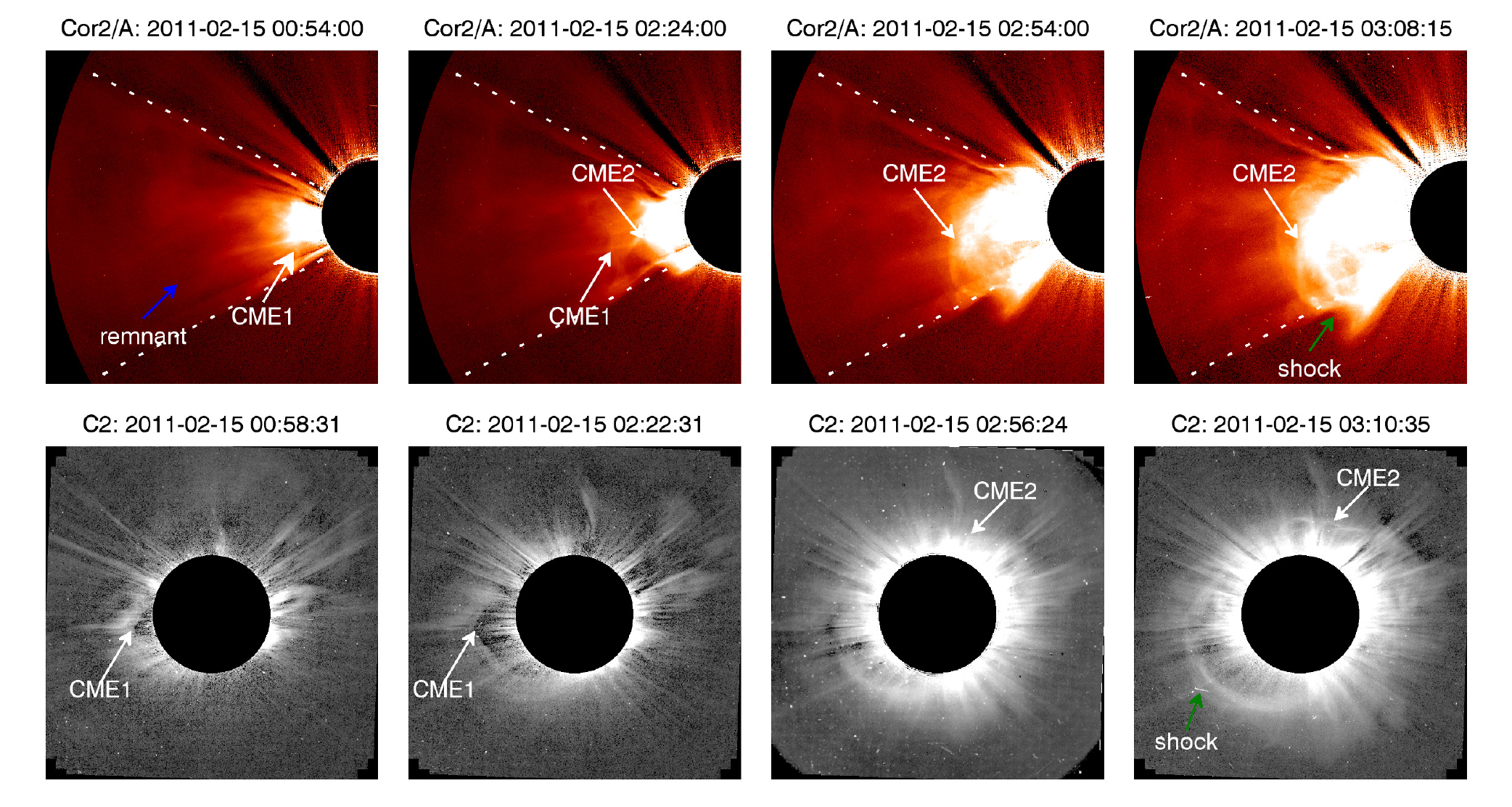}
\caption{Evidence of a preceding CME and a CME-driven shock on 15 February 2011. 
     The preceding CME is marked as CME1 and 
     the CME investigated in this paper is marked as CME2. Upper panels
     show observations from STEREO/COR2/A at different times, lower panels show
     observations from LASCO/C2. The white arrow indicates the projected leading edge of 
     the CMEs. The green arrow indicates the possible shock region. The blue arrow indicates 
     the remnant left from previous CMEs which are widely distributed within the cone region indicated by 
      the two dotted lines in images of STEREO/COR2/A.}
\label{F-precme}
\end{figure*}
\end{center}

F2 also has a relatively high polarization but is oriented azimuthally. 
It is well embedded inside the projection of the halo CME. Because of its 
shape we have to rule out a streamer but our suspicion is that it is due to 
the leading edge of a fairly weak and slow CME (CME1) which was launched about
one hour before the fast and strong halo CME studied here (CME2).
In Figure~\ref{F-precme} we show the propagation of both CMEs. CME1 first appeared in the 
field of view of STEREO/COR2 at 00:54UT. Its propagation speed was estimated to be $\rm 704~km~s^{-1}$ 
along a direction with longitude of $\approx12^\circ$ and latitude of $\approx -6^\circ$ in 
Carrington coordinates using the triangulation of the leading edges observed by STEREO/COR2/A and LASCO/C2 respectively. CME1 was eventually submerged by CME2 
in the both images of STEREO/COR2/A and LASCO/C2 . It is not clear how much the two CMEs interacted,  
but in Figure \ref{F-spectrogram} a radio emission enhancement around 02:40UT between 
the fundamental and its second harmonic could be related 
to such an interaction of the two CMEs\citep{Gopalswamy04}.
    
In the period 13-15 February 2011, a series of CMEs were  
launched successively from the active region AR 11158 \citep{Maricic14}.
A number of them occurred even before CME1 in this time period. One of them launched
at 17:20 UT on 2011 February 14 also interacted with the halo CME studied here. 
This interaction occurred at about 06:49 UT on 2011 February 15 and has been
studied in detail by \citet{Temmer14}. 
From the images of STEREO/COR2 (upper panels in Figure~\ref{F-precme}), 
we can distinguish some remnant from these preceding CMEs which are widely distributed within a cone. 
Their brightness is slightly enhanced in the region between two dotted lines in Figure~\ref{F-precme}.  
Besides the remnant, we can also distinguish a fairly weak CME-driven shock, indicated by the green arrows in Figure~\ref{F-precme}.
The shock was visible at 03:10~UT in the next frame after 02:56~UT when the polarimetric observations of the CME were taken. 
At 02:56~UT, it is very difficult to identify the shock structure. 
Therefore, we suspect that both the remnant and the possible CME-driven shock may contribute to  F3. 
However, considering the weakness of the shock, probably F3 mainly comes from the remnant.  

\begin{center}
\begin{figure*}[!htb]
\centering
\includegraphics[width=0.8\textwidth]{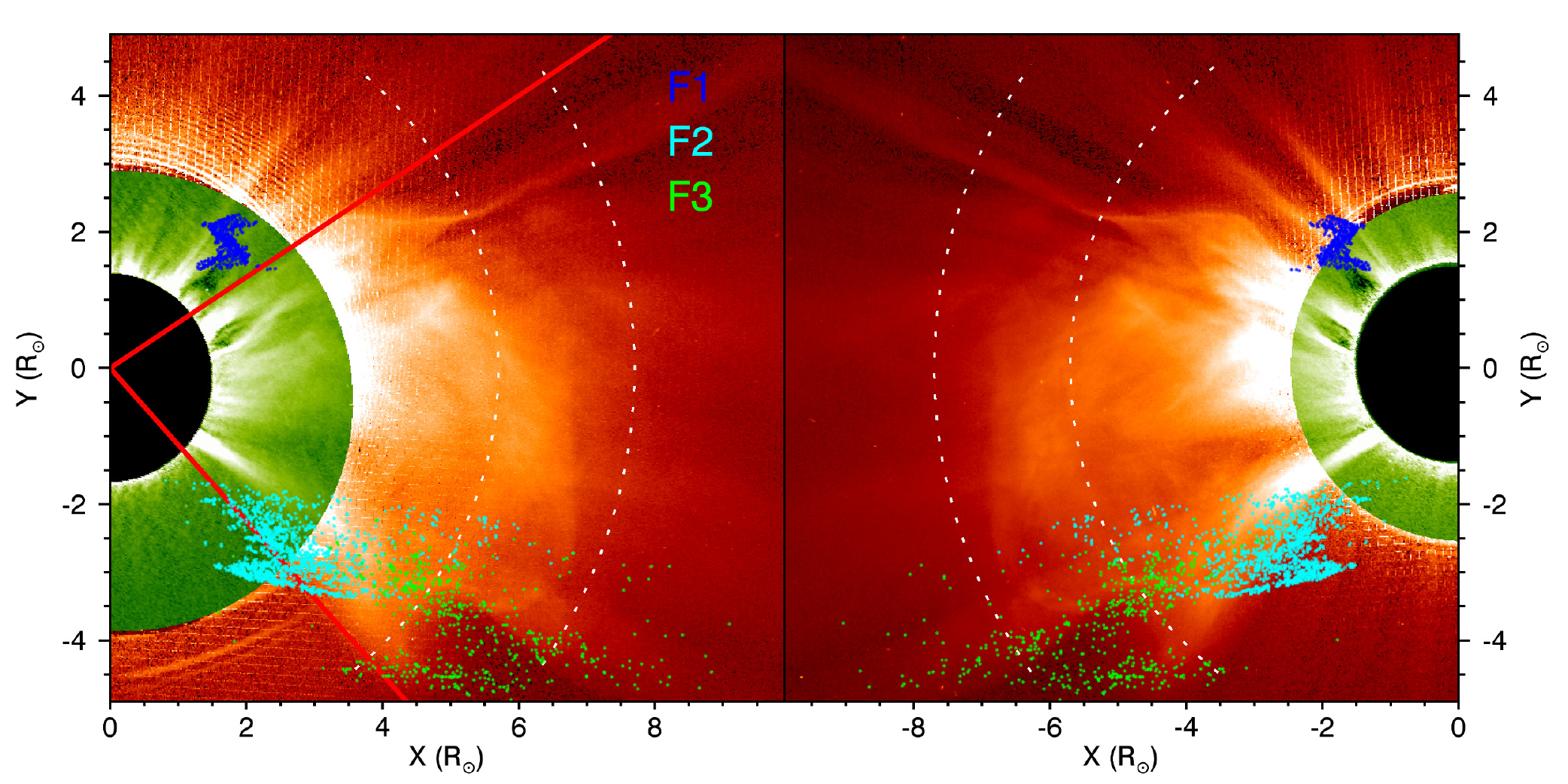}
\caption{The projection of three reconstructed features onto the white light image 
   pair from STEREO A and B. The three features were first distinguished from the 
   image of polarization degree as shown in Figure~\ref{F-features} and then reconstructed by polarimetric method.   
   The red solid line indicates the lower boundary of the cone-shape of the CME. The dotted curves show 
   the uncertainty for the radial distance of the radio source region derived 
   from the radio observations.}
\label{F-featureprojection}
\end{figure*}
\end{center}

The various features derived from LASCO C2 should
match corresponding white-light structures seen from STEREO. 
In Figure~\ref{F-featureprojection} the 3D position of the scattering centers
of different features were projected onto the white-light total brightness images
observed at the same time by STEREO A and B.
As the view direction of LASCO on 15 Feb 2011 is almost
at right angles with the STEREO spacecraft, the $x$-coordinate of the
equivalent scattering centers is almost entirely given by their depth
$z_\circ$ from the PR method.
As a reference, the dotted curves represent the uncertainty for the radial distance of the radio
source region estimated from the radio observations, as discussed in Section~\ref{subsec-rad}. 

\subsubsection{Separation of CME and background in image of polarization degree}\label{subsec-sep}

\begin{center}
\begin{figure*}[!htb]
\centering
\includegraphics[width=0.8\textwidth]{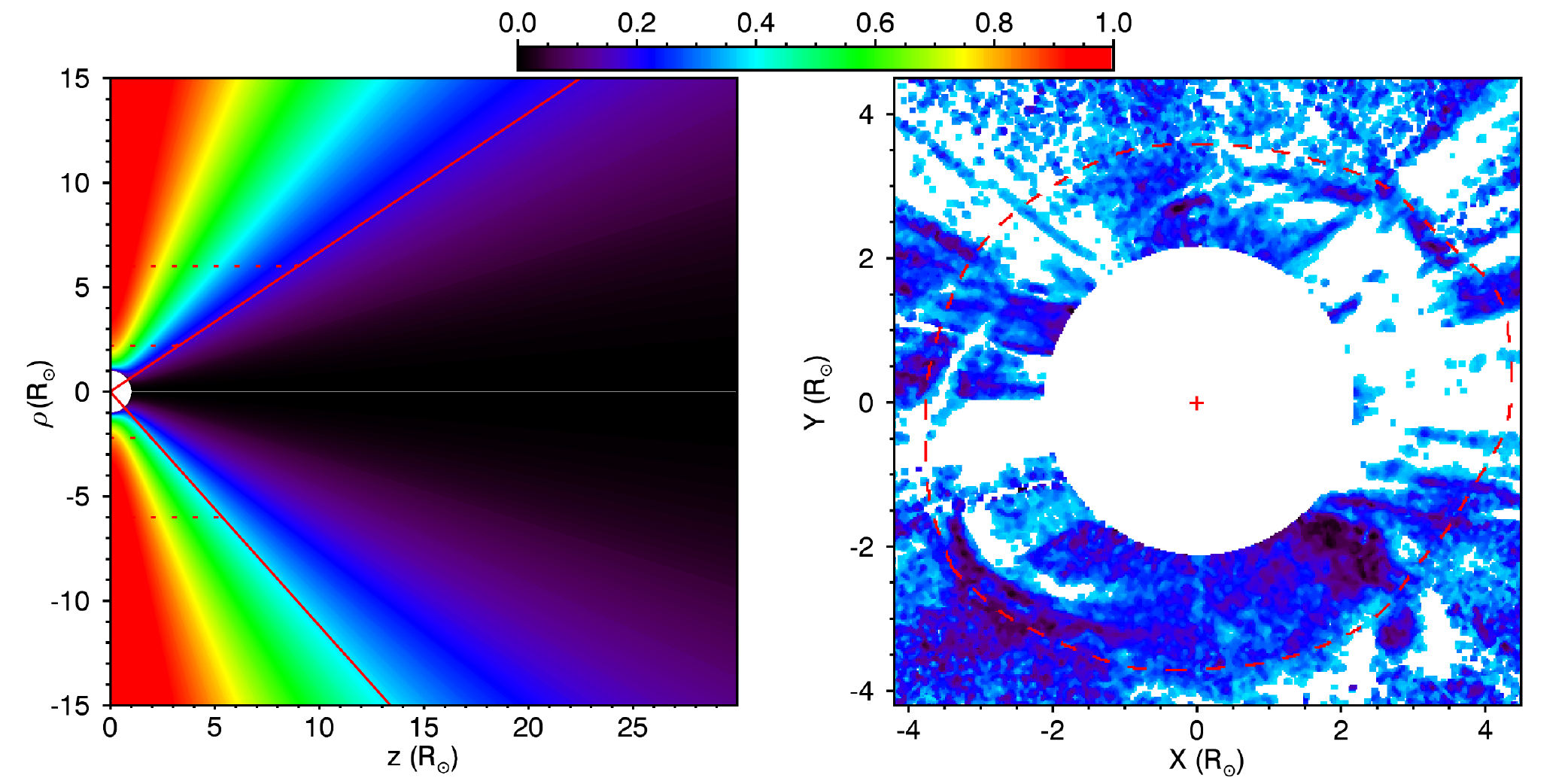}
\caption{The left panel shows the theoretical relationship between  P(polarization degree), 
   z(distance from the POS) and $\rho$ (distance along the POS). The red solid lines are the lateral 
   boundaries of the halo CME distinguished from the image of STEREO/COR2/A, as shown in 
   Figure~\ref{F-featureprojection}. The red parallel dotted lines indicate the field of view of LASCO/C2 ranging from 2.2 to 6 $R_\odot$. 
   The right panel shows the image of polarization degree with background structures (polarization 
   degree $> 0.4$) removed. The red dashed curve indicates the boundary 
    of the CME identified from white-light observations.  The color bar indicates the value of polarization degree.}
\label{F-pdeg}
\end{figure*}
\end{center}

In order to  improve the reliability of 3D reconstruction of the halo CME investigated here 
with the PR method, we attempt to separate the CME features from the background 
structures discussed in Section~\ref{subsec-fea}. According to Thomson Scattering theory, 
the farther away of the electrons from POS, the lower the polarization degree of scattered light. 
The left panel of Figure~\ref{F-pdeg} shows the theoretical relationship between P (polarization degree), 
z (distance from the POS) and $\rho$ (distance along the POS). From the limb views, most CMEs 
can be seen to propagate within a limited angular cone with a half width often of less 45 degrees. 
The boundary of the cone should give an upper limit of the observed polarization degree. It is found 
that the contour lines of constant polarization degree beyond a distance of 1 to 2 $\rm R_\odot$ 
are cone-shaped as well. Therefore, we conclude that the plasma cloud of a well centered halo 
CME (it would propagate outwards along the z direction in the left panel of Figure~\ref{F-pdeg} ) 
should produce coronagraph polarization ratios below about 0.4 once it has propagated more than 
1 to 2 $\rm R_\odot$ from the POS. If this CME propagates self-similarly, the maximum polarization degree would not change much.

For our CME, after varying the threshold of polarization degree in a certain range, 
we found 0.4 is a good value for separating the CME. Such a selection is further verified by 
the cone distinguished from the white-light images of STEREO/COR2 shown in the left panel of 
Figure~\ref{F-featureprojection} by two red solid lines. The CME cone is over-plotted by the red 
solid lines in the left panel of Figure~\ref{F-pdeg}. The FOV of LASCO/C2
 is marked by red parallel dotted lines(the inner boundaries at $\rho=\pm 2.2~ \rm R_\odot$ are caused 
 by the LASCO/C2 occulter). It is obviously that the CME cone 
locates within a contour level of 0.4. We therefore use a polarization degree of 0.4 as a criterion to 
separate CME signals from the background. We show our separating result in the right panel of Figure~\ref{F-pdeg}. 
The image has been enlarged so as to see the details of the cleaned CME. The red dashed curve 
indicates the boundary of the CME identified from white-light observations by LASCO/C2. 
We will use this cleaned image to estimate the 3D structure of the CME.

\begin{center}
\begin{figure*}[!htb]
\centering
\includegraphics[width=0.8\textwidth]{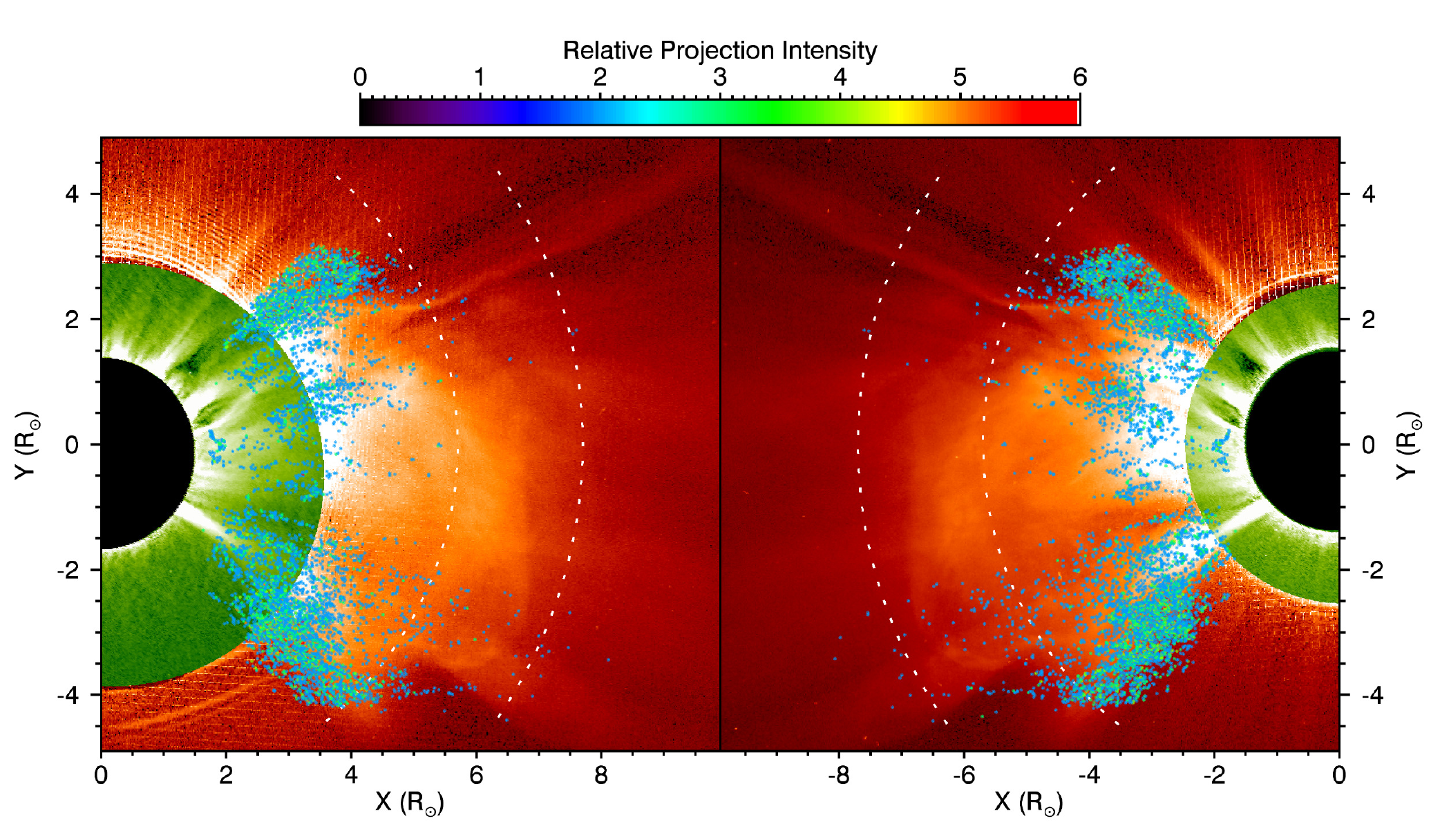}
\caption{A similar figure as shown in Figure~\ref{F-featureprojection}. With this time 
   we projected the reconstructed CME, which was first singled out from background 
   structures as shown in Figure~\ref{F-pdeg}, onto the white light image pair observed by STEREO A and B. 
   The color bar  shows the relative projection intensity.}
\label{F-cmeprojection}
\end{figure*}
\end{center}

\begin{center}
\begin{figure*}[!htb]     
   \centerline{\includegraphics[width=0.8\textwidth,clip=]{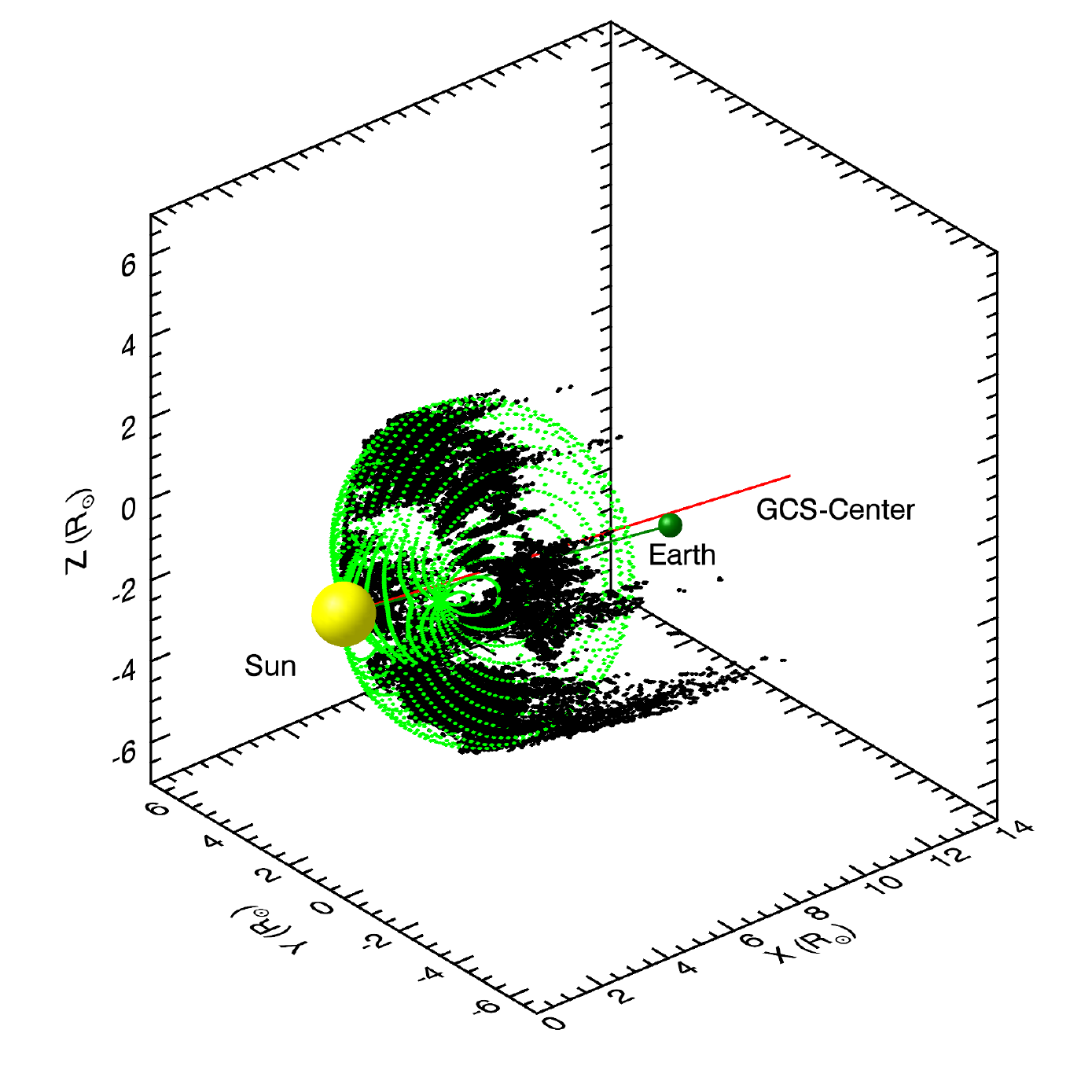} }
   \centerline{\includegraphics[width=0.8\textwidth,clip=]{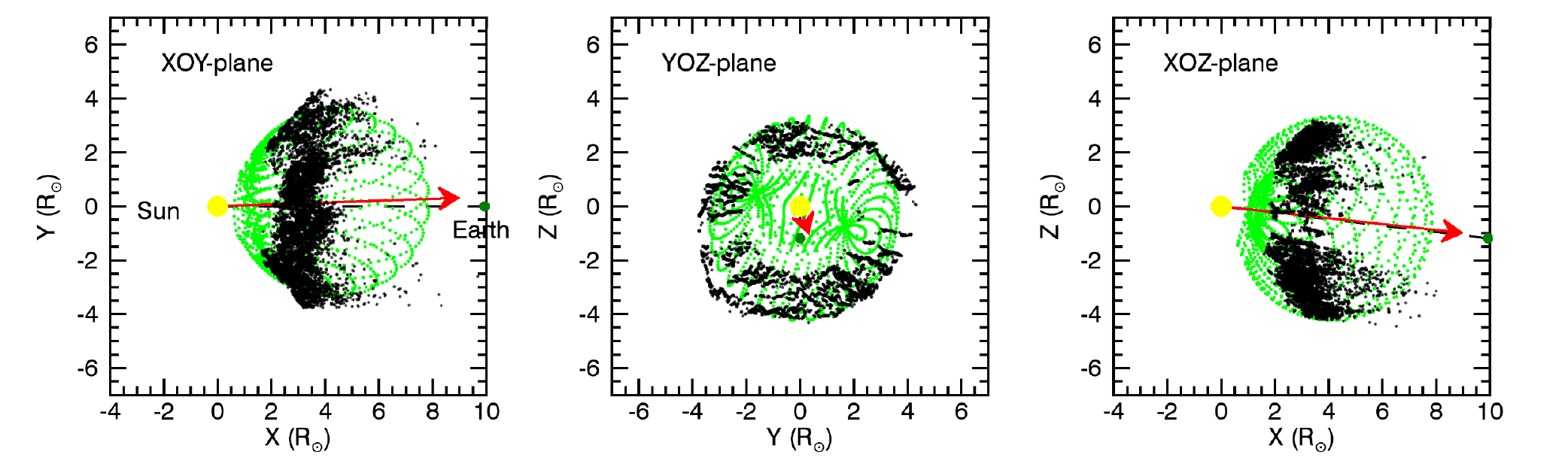} }
   \caption{3D reconstruction of the cleaned halo CME by combining the polarimetric method  with GCS FM. 
   (upper panel) reconstruction result in 3D space, (lower panels) projection of the reconstructed 
   CME onto different 2D planes. The red arrows indicate the direction of the reconstructed CME.}
   \label{F-gcs_pr}
 \end{figure*}
\end{center}

Similar to the projection of the three background features, Figure~\ref{F-cmeprojection} shows the
 projection of the reconstructed CME with background structures removed. The color bar displays the 
 relative projection intensity\footnote{We calculate the relative projection intensity by summing the 
 number of projected 3D CME points in each pixel along the projection direction.}The lack of 
 points at the CME front is due to the occultation 
 of the CME front in the LASCO view, which is an unavoidable problem for coronagraph
 observations of halo CMEs. 
 To give an impression of the entire CME we combine the PR method with the
 GCSFM (GCS-PR) fit procedure applied only to the LASCO observations.
 We first apply  the GCSFM to the LASCO C2 total brightness image to get a first estimate of 
 the free parameters, and then use the 3D points derived from polarimetric method to further 
 constrain the GCSFM parameters.
 Hence the GCSFM results is only based on LASCO halo CME observations. 
 Figure~\ref{F-gcs_pr} shows the fitting result of this method, from which we estimate that the CME 
 has reached a height of $\rm 7.9~R_\odot$ with a direction of longitude $\approx 23^\circ$ 
 and latitude $\approx -6^\circ$ in the carrington coordinate system at the time when the
 LASCO polarization images were
 observed. This height lies at the upper
 limit of the height range determined from the radio frequency measurement and the leading edge
 observed from STEREO (see Figure \ref{F-th}).

\section{Comparison}\label{sec-com}

Figure~\ref{F-th} summarizes the propagation of the projected CME leading edge
triangulated from the stereo observations and
the path of the radio source derived from the frequency drift and the 
interplanetary density model by \cite{Vrsnak04}. The uncertainty of the radio 
source position is indicated by the gray region, and the uncertainties of the leading edge are 
indicated by the vertical bars. The radio source propagation of the type II burst are found 
to be very close to the leading edge of the CME, which complies well with the scenario of 
a CME-driven shock wave. 
The region of radio emission seems to spread out with distance and appears
to move slightly faster than the leading edge of the CME.

Due to the viewing geometry, the positions of the equivalent scattering
centers calculated by the PR method and overplotted into STEREO white-light
images should match respective white-light features in these images.
The scattering center, however, at best represents a mean depth position which
may agree with the real electron density distribution only if it is concentrated in depth.   
Therefore we found it therefore necessary to separate features of different
depth in the polarization degree images and compare them individually.
In Figure~\ref{F-featureprojection} 
we present the projection of three features onto the white-light images of
STEREO/A and B. The
projection of the separated CME is shown in Figure~\ref{F-cmeprojection}. 
We find that the equivalent scattering centers are consistent with the observations from STEREO,
which confirms that the PR method yields reasonable results.

\begin{center}
\begin{figure*}[!htb]
\centering
\includegraphics[width=0.8\textwidth]{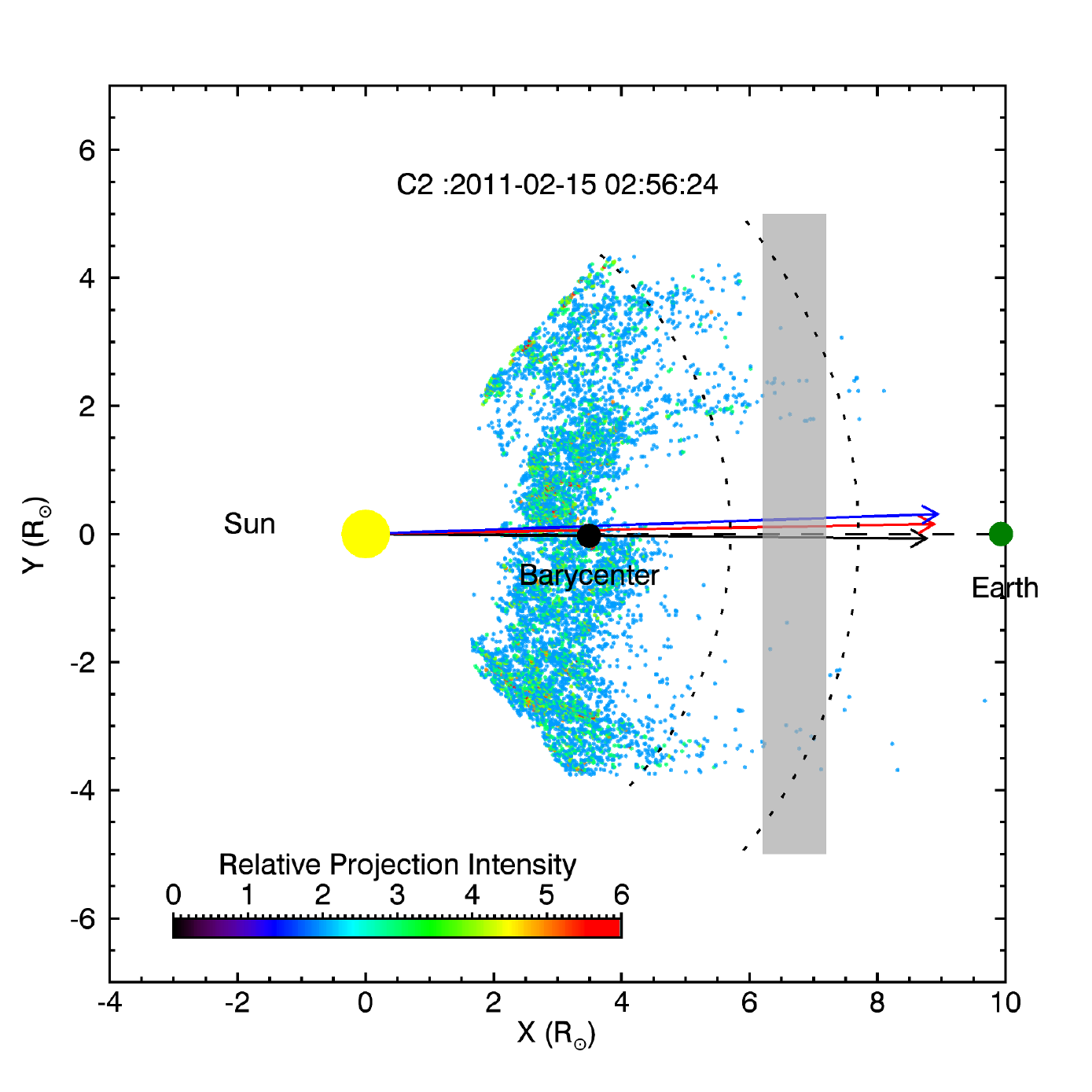}
\caption{Comparisons of the position and direction of the CME at 02:56:24UT derived by different methods. 
   The colorful points represent the projection of the reconstructed CME by PR method onto the solar equator plane. The color 
   bar indicates the relative projection intensity and the black ball represents the projection of the barycenter of the reconstructed CME.
   The dashed line indicates the Sun-Earth connection and the green ball indicates
   the Earth direction. The grey region shows the position range of the CME leading edge as viewed by STEREO A and B. 
   The dotted curves show the uncertainty range for the radial distance of the radio source region estimated from the
   frequency drift observed by Wind/WAVES and the interplanetary density model developed by \citet{Vrsnak04}.
    The black, red and blue arrows represent the direction of the CME barycenter and the directions of the CME derived 
    from GCS-PR and GCS-Tri, respectively.}
\label{F-comparison}
\end{figure*}
\end{center}

The line-of-sight integration effect is even more pronounced for the halo CME.
The CME is a distributed density cloud, and we cannot expect that the
distribution of the scattering centers we attributed to the CME
exhibit its entire shape. 
For a quantitative comparison, we combine these scattering centers to their barycenter  
and compare it to the GCSFM fitting results and to the location of the leading edge 
derived from STEREO observations.
This comparison is shown in Figure~\ref{F-comparison}.
The grey region indicates the distance range of the CME leading edge from
the STEREO observations and the two dotted lines 
represent the possible boundaries of the radio source region at the
time when the polarized images were taken by LASCO.
The  barycenter of the equivalent scattering centers was found at
a distance of $\rm 3.7~R_\odot$ from the Sun center, about $\rm 3~R_\odot$ behind the leading
edge at the time when the LASCO polarized images were taken.
This discrepancy is for once due to the line-of-sight integration effect which
causes the scattering centers to represent the entire CME density rather than
the leading edge. In addition, the occultation of the central CME region from
the LASCO observations may also play a role. 
In Figure~\ref{F-cmeprojection}, where we have overplotted the equivalent
scattering centers attributed to the CME on top of the STEREO images it is
obvious that some central and most advanced parts of the CME cloud are not
covered by equivalent scattering centers.

The directions of the CME estimated from both GCSFM to image triplets (GCS-Tri)
and GCSPFM to PR results(GCS-PR) are very close to the CME barycenter
direction. A comparison between GCS-Tri and GCS-PR is shown in Table \ref{tab-cme}.
The difference seems very small, which indicates that a combination of GCSFM with
PR method could be a feasible approach to measure the propagation direction
and distance of a halo CME from a single perspective at Earth.


 \begin{table}[!htb]
\centering
\caption {Longitude and latitude of reconstructed CME in the Carrington coordinate
 system on 15 February 2011.}
\begin{tabular}{ccc}
\hline \hline
 & GCS-PR & GCS-Tri  \\
\hline
 Longitude($^\circ$) & 23.14 & 22.14\\
 Latitude($^\circ$) & -6.30 & -9.30 \\
  Height($\rm R_\odot$) & 7.90 & 7.10 \\
\hline
\hline
\end{tabular}
\label{tab-cme}
\end{table}

\section{Summary} \label{sec-sum}      
  The goal of this study was to explore the possibilities to measure the 
  approach of a halo CME to Earth by observations from the Earth-direction 
  alone. In the not too far future, when STEREO will not be available any 
  more, these observations will be the only data on which a CME warning 
  could be built upon. We focused in particular two methods to achieve 
  this goal: type II radio burst inversion and polarization ratio method. For this 
  study, we select a full halo CME on 15 February.
  On that day, STEREO A and B each had a separation angle of about 
  $90^\circ$ with SOHO, which means the halo CME moved close to the POS
  with respect to STEREO A and B. Firstly, this allowed to directly and
  reliably record the propagation of the CME leading edge as a reference for
  the results of the other two methods.
  Secondly, the favorable viewing geometry of the
  three spacecraft allowed to compare the depth estimates derived from polarized 
  images taken by LASCO directly with corresponding white-light features in
  the STEREO image set.
   
   Unfortunately, LASCO only produced a single polarimetric image set of the
   halo CME at about 02:54 so that we could compare the results of the PR
   method only for this single instance.
   Coronagraph images of a halo CME are heavily contaminated by different background 
   structures as seen, for example, in Figure~\ref{F-features}.
   Using images of the polarization ratio can greatly help to disentangle the
   various structures and separate the CME signals from the background. 
   In our case, we combine the CME cone observed by STEREO/COR2 with the 
   theoretical model of polarization degree to obtain a criterion value(polarization degree 
   = 0.4) for separating the CME signals.
   The result of this separation was demonstrated in the right panel of Figure~\ref{F-pdeg}.
   In future, for the case of lacking STEREO observations, we maybe apply the typical 
    cone model to get an estimate of the criterion of polarization degree\citep{Zhao02,Xie04}.

   The comparison of the depth estimates from the PR method with white-light features in
   the STEREO images shows good consistency if the comparison is made for
   individual features identified in the polarization ratio images by a
   locally uniform polarization ratio. The more diffuse CME cloud, on the
   other hand has a large depth extent while the PR method yields only a
   single representative depth estimate for each pixel analyzed.
   A relationship of the position of the equivalent scattering centers with the
   CME leading edge is not easily established. The maximum distance of the
   scattering centers from the Sun is a statistically very noisy measure. 
   The barycenter of the equivalent scatterers, even when cleaned from
   background features, gave a distance of about half of the leading edge
   distance. Further studies like the one presented here might eventually yield
   a well defined correction factor between the barycenter and the leading
   edge distance. We found it helpful
   to combine the PR method with the GCSFM fit to derive the probable shape of
   the CME cloud which gave a better match for the leading edge distance
   estimate and also for the direction of the CME propagation.
   
   The second method to analyze the time-frequency variation of the type II
   radio burst associated with the CME also has some uncertainties. The most
   critical assumption is the interplanetary density model required to convert
   plasma frequency into distance from Sun center. We found the model developed
   by \cite{Vrsnak04} gave the best agreement to the kinematics of the leading
   edge as derived from the STEREO observations. This comparison is summarized
   in Figure~\ref{F-th}.  The discrepancy in the estimated velocities
   amounted to about 5\% with the radio burst frequency giving a slightly
   faster speed estimate. Extensions of the method used here also analyze the
   correlation of the radio burst signal in different antenna polarizations
   from which the direction and the size of the radio source can be estimated
   \citep{Manning80,Cecconi05}.
   
   A problem with the radio frequency method for routine CME arrival
   predictions is that not all CMEs emit a well identifiable signal. For that
   reason, the PR method, even though it yields less precise results should
   be considered as a back-up when the radio signals are too weak or are
   missing completely. We hope that an extension of the study presented
   in this paper to more halo CME events will help to improve the analysis of
   the PR method and eventually lead to more precise CME predictions.
  
  \section*{Acknowledgments}  
 We acknowledge the use of data from STEREO, SOHO and WIND. 
 Many thanks to Thomas G. Moran and Angelos Vourlidas for their helpful discussions on dealing 
 with LASCO polarimetric data. This work was supported  by the NSFC grants
 (11522328, 11473070, 11427803, 11233008 and 11273065)  and by the Strategic Priority 
 Research Program, the Emergence of Cosmological Structures, of the Chinese Academy of
Sciences, Grant No. XDB09000000.
L.F. also acknowledges the Youth Innovation Promotion Association and the specialized research fund from State Key Laboratory of Space Weather for financial support. 
               
\bibliographystyle{aasjournal}
\bibliography{work-3D}

\end{document}